\documentclass[11pt,a4paper]{article}
\usepackage{amsmath}
\usepackage{setspace}
\usepackage{xypic}
\usepackage[all]{xy}
\usepackage{latexsym}
\usepackage{theorem}
\newtheorem{teorema}{Theorem}[section]
\newtheorem{definicion}[teorema]{Definition}
\include{amslatex}
\newtheorem{proposicion}[teorema]{Proposition}

\newtheorem{corolario}[teorema]{Corollary}

{\theorembodyfont{\rmfamily}
}
{\theorembodyfont{\rmfamily}
}

\numberwithin{equation}{section}

\include{amsmath}

\begin{document}

\begin{title}
{\LARGE {\bf Fluid Models from Kinetic Models using a Geometric Averaging Procedure}}
\end{title}
\maketitle
\author{
\paragraph{}
\paragraph{}
\begin{center}

Ricardo Gallego Torrome\\
Department of Physics, Lancaster University,\\
Lancaster, LA1 4YB \& The Cockcroft Institute, UK\footnote{email:
 r.gallegotorrome@lancaster.ac.uk ; Partially
supported by EPSRC.}\\ [3pt]
\end{center}}

\bigskip
\begin{abstract}
We interpret the Lorentz force equation as a geodesic equation
associated with a non-linear connection. Using a geometric
averaging procedure, we prove that for narrow and smooth
 one-particle distribution functions whose supports are invariant under the flow of the Lorentz equation,
  a bunch of charged point particles can be described by a charged cold fluid model in the ultra-relativistic
 regime. The method used
 to prove this result does not require additional hypotheses on the
higher moments of the distribution. This is accomplished by estimating
    the expressions that include the differential operators appearing in the
     charged cold fluid model equation.
    Under the specified conditions of narrowness and ultra-relativistic dynamics,
    it turns out that these differential expressions are close to zero, justifying
    the use of the charged cold fluid model. The method presented in the work can also be applied to justify the use
     of warm plasmas and other models.
\end{abstract}
\paragraph{}

\section{Introduction}

Despite limitations concerning the mathematical description of the discrete
nature of the particles comprising a plasma,
modeling the dynamics of relativistic non-neutral plasmas and charged particles beams
by fluid models is a common place. The relative simplicity of these models (compared with the
 corresponding kinetic models) is a partial justification of their use. We propose in this work another justification. We will concentrate in the charged cold fluid model. However, we should notice that the same philosophy is applicable to more sophisticated models.

In high intensity beam accelerator machines, each bunch in a
 beam contains a large number of identical particles contained in a small
  phase-space region. In such conditions,
 a number of the order of $10^9-10^{11}$ charged particles move {\it together} under
   the action of both external and internal electromagnetic fields. Often in modern applications, such beams of particles are ultra-relativistic.

One is interested in modeling these physical systems in such a way that:
\begin{enumerate}
\item  The model for a bunch of particles must be {\it simple}, in order to be useful in numerical simulations of beam dynamics and for analytical treatment,

\item  It allows for stability analysis and a qualitative understanding of the dynamical behavior of the
 system. Three dimensional numerical simulations can be also desirable.

\end{enumerate}

 The standard approach has been to use fluid models as an approximation to a kinetic model.
 These derivations of fluid models from kinetic models can be found in [1-4] and references there.
They are based on some assumptions, usually in the form of equations of
  state for the fluid or assumptions on the higher moments of the distribution function.
  This constraint is necessary to close the hierarchy of moments of
   the distribution function and to have a sufficient number of differential relations
    to determine the remaining moments. This is a general feature of all the derivations of
    fluid models from Kinetic Theory (see for instance [1-4] and references there),
    where a truncation scheme is required for the fluid model to be predictive.

We present in this work a new {\it justification} of the charged cold fluid model from the framework
 of kinetic theory. The novelty of the new approach is that it uses {\it natural hypotheses} suitable for particle
  accelerator machines and exploit only the mathematical structure of the
   classical electrodynamics of point particles interacting with external electromagnetic fields. In particular, we estimate the covariant derivative of
   the mean velocity calculated with the one-particle distribution function. This is given as an
    asymptotic formula in terms of the time of evolution, diameter of the distribution and energy of the beam.
The fluid model is described by only one {\it dynamical} variable, the
  normalized mean velocity field. The variance and the heat flow tensor are not necessarily zero but
  are finite. In our treatment both the fluid energy tensor and the fluxes tensor are non-dynamical.
  This is why we have considered here the charged cold fluid model. However, the approach is
  consistent, since our aim is not to give an equation for the mean velocity field, but evaluate
   how much certain differential expressions (formally equivalent to the charged cold fluid model equations)
   differ from zero. One can stipulate the validity of the model from the estimate of the
    corresponding differential expressions. On the other hand, in the models presented for instance in
     [1-4], the variance and the covariant heat flow are
     dynamical and a system of partial differential equations is used to
     determine the dynamics of these fields. However, assumptions must be done to close the system of
     fields equations and the validity of those assumptions is unclear.

 The method used to obtain these bounds is the following:
  \begin{enumerate}
 \item Firstly, we re-write the Lorentz force equation as an auto-parallel condition of a non-affine linear connection.

 \item We use an averaging procedure described in [5] to average this connection.

 \item The resulting averaged connection is an affine connection on the manifold {\bf M}.

 \item  It happens that under the same assumptions as used for the particle dynamics [6], the
 corresponding solutions of the Vlasov equation $f$ and the averaged Vlasov equation $\tilde{f}$ are similar. This comparison result is based on the comparison results of the point dynamics.

 \item Given the distributions $f$ and $\tilde{f}$, one has the corresponding mean velocity vector fields. One can prove under the same hypotheses that these mean velocity vector fields are similar, which means that the difference between them is controlled by powers of a small parameter.

 \item Finally, we show here that the auto-parallel condition of velocity field of the averaged Vlasov equation associated
  with the averaged dynamics is controlled by the diameter of the distribution $f$.  Together with the above point, this result provides  estimates for the auto-parallel condition of the mean velocity field of the solution of the Vlasov equation.
 \end{enumerate}

 Therefore, the methods presented here and the usual derivations of the fluid models contained in [1-4]
 are different. The standard approaches assume an asymptotic expansions of the differential equations
 for the moments, in terms of a perturbation parameter which is similar to the diameter $\alpha$ of
 the distribution function. These asymptotic expansions are realized at low orders as a {\it truncation scheme}
 in hierarchy of moments. Then, they discuss systems of partial differential equations such that they are
 self-contained and consistent with physical constraints and with the asymptotic expansions.
  On the other hand, our approach is based on the structure of the Lorentz force equation of a charged
  point particle, which lies at the basis of the kinetic models. After being written in a geometric way, the Lorentz force equation is substituted by the averaged Lorentz force
    equation. The key points are that the averaged Lorentz force equation depends only on the first,
    second and third moments of the distribution function and that is a simpler equation than the original Lorentz force equation.
    Then, under some smoothness assumptions on the distribution function, we can place bounds on the
     differential expression of interest.

There are some assumptions that we have used in the present work. For instance,
 we assume that the distribution functions have compact support. This excludes important examples like Gaussian distributions, which are the distributions in equilibrium. However, one can consider truncated Gaussian distribution as approximations, when the truncation is performed smoothly, using a bumpy functions techniques.

 Also we have assumed that the distributions functions are smooth (at least of class $\mathcal{C}^1$).
 Although we do not currently have a proof that we can extent our results to bigger
functional spaces for the distribution functions, since the main results are written in terms of Sobolev norms, it is conjectured that they can be extended to Sobolev spaces.

{\bf Notation}.

Let $(x,{\bf U})$ be a local coordinate system on ${\bf M}$ ,
where ${\bf U}\subset {\bf M}$ is an open sub-set and $x:{\bf
U}\rightarrow {\bf R}^n$ a local coordinate system. An arbitrary
tangent vector at the point $p\in {\bf U}$ is of the form
$X_p:=X=X^k\frac{\partial}{\partial x^k}|_p$. The local
coordinates associated to the tangent vector $X_p\in {\bf T}_x{\bf
M}\subset{\bf TM}$ are $(x^k,y^k)$. {\bf M} will be $n$-dimensional and equipped with a metric of signature $(+,-,...,-)$. We call ${\bf M}$ space-time manifold. In some calculations it will be useful to consider the particular case when the metric $\eta$ is flat. Given a $1$- form $\omega$, ($\omega^\sharp$ means duality defined by the semi-Riemannian metric $\eta$.

There are certain sub-bundles of the tangent bundle that are relevant in our treatment:
\begin{enumerate}

\item The slit tangent bundle  ${\bf N}:=\bigsqcup_{x\in {\bf TM}}\, \{y\in\, {\bf T}_x{\bf M},\,\,\eta_{ij}(x)\,y^iy^j>0\,\}$,

\item The unit hyperboloid bundle  ${\bf \Sigma}:=\bigsqcup_{x\in {\bf TM}}\, \{y\in\, {\bf T}_x{\bf M},\,\,\eta_{ij}(x)\,y^iy^j=1\,\}$

\item The null bundle ${\bf NC}:=\bigsqcup_{x\in {\bf TM}}\, \{y\in\, {\bf T}_x{\bf M},\,\,\eta_{ij}(x)\,y^iy^j=0\,\}$.
\end{enumerate}
The particularization to each point $x\in {\bf M}$ are $N_x$ and $\Sigma_x$ respectively.

We are dealing with one-particle distribution function $f(x,y)$. We assume that $f_x:=f(x,\dot)$ has compact support on the unit hyperboloid bundle ${\bf
\Sigma}$. The diameter of the distribution $f_x$ is
$\alpha_x:=sup\{d_{\bar{\eta}}(y_1,{y}_2)\, |\, y_1,{y}_2\in
supp(f_x )\}$. Then, we define
${\alpha}:=sup\{{\alpha}_x,\, x\in {\bf M}\}$. We define the energy function $E$ of a distribution $f$ to be the real function:
\begin{displaymath}
E:{\bf M}\longrightarrow {\bf R}
\end{displaymath}
\begin{displaymath}
x\mapsto E(x):=inf\{ y^0,\, y\in supp(f_x)\},
\end{displaymath}
where $y^0$ is the $0$-component of a tangent vector of a possible trajectory of a point particle,
 measured in the laboratory coordinate frame.

 We introduce from reference [6] our notion of semi-Randers space,
\begin{definicion}
A semi-Randers space consists of a triplet $({\bf M},\eta
,[A])$, where ${\bf M}$ is a space-time manifold, $\eta$ is a semi-Riemannian
 metric continuous on {\bf M} and it is smooth on ${\bf TM}\setminus {\bf NC}$ and
  the class of locally smooth $1$-forms $A\in [A]$ defined such that ${\bf F}=dA$ for any $A\in {\bf F}$.
\end{definicion}

\section{Lorentz Force Equation and Averaged Lorentz Force Equation}

We proposed in [6] a geometric description of the dynamics of a charged point
particle interacting with an external electromagnetic field. The relevant data was extracted
 from a Lorentzian metric $\eta$ with signature $(+,-,-,-)$ and from the Lorentz force equation,
  that in an arbitrary local coordinate system reads [7,8]:
\begin{equation}
\frac{d^2 \sigma^i}{d\tau^2} +\, ^{\eta}\Gamma^i\,_{jk} \frac{d
\sigma^j}{d\tau}\frac{d \sigma^k}{d\tau} +\eta^{ij}(dA)_{jk} \frac{d
\sigma^k}{d\tau}\sqrt{\eta(\frac{d \sigma}{d\tau},\frac{d
\sigma}{d\tau})}=0,\quad i,j,k=0,1,2,3,
\end{equation}
where $\sigma: {\bf I}\longrightarrow {\bf M}$ is a solution, a curve parameterized by
 the proper-time $\tau$ associated with $\eta$, $^{\eta}\Gamma^i\,_{jk}$ are
 the coefficients of the Levi-Civita connection $^{\eta}\nabla$ of $\eta$ and $dA={\bf F}$ is
  the exterior derivative of the $1$-form $A$.

We viewed the system of differential equations $(1.1)$ as the auto-parallel condition of a connection
that we called the Lorentz connection $^LD$. In reference [6] were defined several related connections: the non-linear connection
  [9,10], the linear connection in the sense of Koszul [11] on a suitable sub-bundle of ${\bf TTM}$ and an example of a connection in
  the pull-back bundle $\pi^*{\bf TM}$ [7]. These connections are related. For instance,
   the non-linear connection
   determines the linear connection in the sense of Koszul. The non-linear
   connection (with some additional assumptions on the {\it torsion-type} tensor) also determines
    the linear connection on the pull-back bundle. In this work we consider
    mainly Koszul connections (which we will denote by $^LD$ and $<\,^LD>$) as determined
    in reference [6], since this allows us to introduce covariant derivatives and auto-parallel equations.

Let us denote by $\eta(Z,Y):=\eta_{ij}(x)Z^i\, Y^j$,
\begin{definicion}
For each tangent vector $y\in {\bf T}_x{\bf M}$ with $\eta(y,y)>0$, there are defined the following functions:
\begin{displaymath}
^L\Gamma^i\,_{jk}(x,y)=\, ^{\eta}\Gamma^i\,_{jk} +
\frac{1}{2{\sqrt{\eta(y,y)}}}({\bf F}^i\,_{j}(x)y^m\eta_{mk}+ {\bf
F}^i\,_{k}(x)y^m\eta_{mj})+
\end{displaymath}
\begin{equation} +{\bf F}^i\,_m (x)
\frac{y^m}{2\sqrt{\eta(y,y)}}(\eta_{jk}-\frac{1}{\eta(y,y)}\eta_{js}
\eta_{kl}y^s y^l),
\end{equation}
where $\eta(y,y)$ is an abbreviation of $\eta_{ij}(x)y^iy^j$, $^{\eta}{\Gamma}^i\,_{jk},\,
(i,j,k= 0,1,2,...,n)$ are the connection coefficients of the Levi-Civita connection
$^{\eta}\nabla$ in a local frame, ${\bf F}_{ij}:=\partial_i A_j -\partial_j A_i$ and ${\bf F}^i\,_j=\eta^{ik}{\bf F}_{kj}$.
\end{definicion}
\begin{proposicion} The Lorentz force equation can be written as the auto-parallel condition
\begin{displaymath}
^LD _{\dot{\tilde{x}}} \dot{\tilde{x}}=0,
\end{displaymath}
where ${x}:{\bf I}\longrightarrow {\bf M}$ is a time-like curve parameterized
 with respect to the proper time of the Lorentzian metric $\eta$,
$\tilde{x}$ is the horizontal lift on {\bf N} and $^LD$ is the
Koszul linear connection determined with connection coefficients given by equation $(2.2)$.
\end{proposicion}
{\bf Proof}: One can check this fact by direct computation (for instance see [6]).\hfill$\Box$

We will define averaging procedure for several geometric objects. These averaging operations
 correspond to {\it fiber integrations} of sub-bundles of {\bf TM}.
  (in particular we are interested in objects living on the unit hyperboloid sub-bundle).
Usually the measure is given as $f(x,y) dvol(x,y)$, where $f(x,y)$ is the one-particle
probability distribution function [12]. The volume form $dvol(x,y)$ is induced by the Lorentzian metric $\eta$:
\begin{displaymath}
dvol (x,y)=\sqrt{-det\,{\eta}}\,\frac{1}{y^0}\,dy^1\wedge\cdot\cdot\cdot dy^{n-1}.
\end{displaymath}
One can prove the following
\begin{proposicion}
The averaged connection of the Lorentz connection $^LD$ defined on the
pull-back bundle $\pi^*{\bf TM}\rightarrow {\bf N}$ is an affine,
symmetric connection on ${\bf M}$. The connection coefficients are given by the formula:
\begin{displaymath}
<\,^L\Gamma^i\,_{jk} >=\, ^{\eta}\Gamma^i\,_{jk}+ ({\bf
F}^i\,_{j}<\frac{1}{2\sqrt{\eta(y,y)}}y^m>\eta_{mk}+ {\bf
F}^i\,_{k}<\frac{1}{2\sqrt{\eta(y,y)}}y^m>\eta_{mj})+
\end{displaymath}
\begin{equation}
+{\bf F}^i\, _m\,\frac{1}{2}\big(
<\frac{y^m}{({\eta(y,y)})^{3/2}}>\,\eta_{jk}-\eta_{js}
\eta_{kl}<\frac{1}{(\eta(y,y))^{3/2}}\,y^m y^s y^l>\,\big).
\end{equation}
Each of the integrations is equal to the $y$-integration along the
fiber:
\begin{displaymath}
vol({\bf \Sigma}_x):=\int _{{\bf \Sigma}_x}
f(x,y)\,dvol(x,y),\quad <y^i> :=\frac{1}{vol({\bf \Sigma}_x)}\int _{{\bf \Sigma}_x} y^i
f(x,y)\,dvol(x,y),
\end{displaymath}
\begin{displaymath}
<y^i y^j>:=\frac{1}{vol({\bf \Sigma}_x)}\int_{{\bf \Sigma}_x}\,y^i y^j\, f(x,y)\,dvol(x,y),
\end{displaymath}
\begin{displaymath}
<y^i y^j y^k>:=\frac{1}{vol({\bf \Sigma}_x)}\int_{{\bf \Sigma}_x}\,y^i y^j y^k\,f(x,y)\,dvol(x,y).
\end{displaymath}
\end{proposicion}
\begin{definicion}
There is a Koszul connection defined on {\bf M}
 such that given sections $X,Y\in \Gamma{\bf T M}$ is defined by:
\begin{equation}
<\,^LD>_X Y:=\,\big(X^j\,\frac{\partial Y^i}{\partial x^j}\,+<\,^L\Gamma^i\,_{jk}>(x)\,X^i Y^j\big)\,\partial_i,\quad i,j,k=0,\cdot\cdot\cdot, n-1.
\end{equation}
We call this connection the averaged Lorentz connection.
\end{definicion}
\begin{proposicion}
Let ${\bf M}$ be the space-time manifold and $A$ a $1$-form on {\bf M}. Assume the existence of a non-negative
function $f:{\bf \Sigma}\longrightarrow {\bf R}$ with compact support. Then
\begin{enumerate}

\item The connection $<\,^LD>$ is an affine, symmetric connection on ${\bf
M}$. Therefore, for any point $x\in {\bf M}$, there is a {\it normal coordinate system} such that the averaged connection coefficients are
zero.

\item  $<\,^LD>$ is determined by the first, second and third moments of the distribution function $f(x,y)$.
\end{enumerate}
\end{proposicion}

Later we will obtain bounds on the value of certain covariant derivatives. These bounds are obtained using a
norm, defined in a particular local frame. This local frame is associated with the {\it bulk motion of the fluid}.
In particular the local frame is determined using the mean velocity field $<y^i>\frac{\partial}{\partial x^i}$.

In a Lorentzian manifold $({\bf M},\eta)$ where there is defined on ${\bf  M  }$ a time-like vector field $U$ normalized such that
$\eta(U,U)=1$, one can define the Riemannian metric $\bar{\eta}$ :
\begin{equation}
\bar{\eta}(X,Y):=-\eta(X,Y)+2\eta(X,U)\eta(Y,U).
\end{equation}
We choose the following locally smooth vector field $U$ in the definition of the Riemannian metric $(2.5)$:
for $\eta_{ij}(x)<\hat{y}^i>(x)<\hat{y}^j>(x)\,> 0$, $U(x)$ is given by
\begin{equation}
U(x)=\frac{<\hat{y}>(x)}{\sqrt{\eta_{ij}(x)\,<{y}^i>(x)<{y}^j>(x)}},
\end{equation}
This vector field is not continuous, since there is a discontinuity in the boundary $\partial(\pi(supp(f))$. Also let us note that for at each $x$ and for functions $f$ with convex supports on the connected hyperboloid, the mean velocity field is time-like.
From now, we will assume that the support of the distribution function is connected.

Using $\bar{\eta}$ there is a Riemannian metric on each
vector space ${\bf T}_x{\bf M}$ defined as
$\bar{\eta}_{ij}(x) dy^i\otimes dy^j$. It induces a distance
function $d_{\bar{\eta}}$ on the manifold ${\bf T}_x{\bf M}$ and related operator norm [13];
given a continuous operator $O_x:{\bf T}_x{\bf
M}\longrightarrow {\bf T}_x{\bf M}$, its operator norm is defined by
\begin{displaymath}
\|O_x\|_{\bar{\eta}}:=sup\Big\{\,\frac{\|O_x(y)\|_{\bar{\eta}}}{\|y\|_{\bar{\eta}}} (x),\,y\in{\bf T}_x{\bf M}\setminus \{0\}\,\Big\}.
\end{displaymath}

Let us restrict to the case where the Lorentzian metric is the Minkowski metric in dimension $n$.
Let us denote by $\bar{\gamma}(t)$ the gamma factor of the Lorentz transformation
from the local frame defined by the vector field $U$ to the laboratory frame, at
 some instance defined by the local time $t$, the coordinate time defined by the
  laboratory frame. Denote by $\theta ^2(t)=\vec{y}^2(t)-\,<\vec{\hat{y}}>^2(t)$
 and $\bar{\theta}^2(t)=<\vec{\hat{y}}>^2(t)-\vec{\tilde{y}}^2(t)$. Here $\vec{y}(t)$
   is the velocity tangent vector field along a solution of the Lorentz force equation
    and $\tilde{y}(t)$ is spatial component of the tangent vector field along a solution
     of the averaged equation, with both solutions having the same initial conditions.
      The maximal values of this quantities on the compact space-time manifold are denoted
       by $\theta^2$ and $\bar{\theta}^2$.

 The basic result from [6] that we use here is
\begin{teorema}
Let $({\bf M},\eta, [A])$ be a semi-Randers space and $\eta$ the Minkowski metric. Let us assume that:
\begin{enumerate}

\item The auto-parallel curves of unit velocity of the connections $^L\nabla$ and
 $<\,^L\nabla>$ are defined for the time $t$, the time coordinate measured in the laboratory frame.

\item The dynamics occurs in the ultra-relativistic limit, $E(x)>>1$ for all $x\in {\bf M}$.

\item The distribution function is narrow in the sense that ${\alpha}<<1$ for all $x\in {\bf M}$.

\item It holds the following inequality holds:
\begin{displaymath}
|\theta^2\,-\bar{\theta}^2|\ll 1,
\end{displaymath}
\item The support of the distribution function $f$ is invariant under the flow of the Lorentz force equation.

\item The following adiabatic condition holds: $\frac{d log E}{dt}<<1$.
\end{enumerate}
Then, for the same arbitrary initial condition $(x(0),\dot{x}(0))$, the solutions of the equations
\begin{displaymath}
^L\nabla_{\dot{x}} \dot{x}=0,\, \quad <\,^L\nabla>_{\dot{\tilde{x}}} \dot{\tilde{x}}=0
\end{displaymath}
 are such that:
\begin{equation}
\|\tilde{x}(t)-\, x(t)\|\leq\, 2\big(C(x)\|{\bf F}\|(x)\,+C^2_2(x)(1+B_2(x){\alpha})\big)
{\alpha}^2\,E^{-2}(x)\,t^2,
\end{equation}
where the functions $C(x)$, $C_i(x)$ and $B_i(x)$ are bounded by constants of order $1$.
\end{teorema}
In a similar way, one can compare the velocity tangent fields along the corresponding geodesics [6]:
\begin{teorema} Under the same hypothesis as in {\it theorem
2.6}, the difference between the tangent vectors is given by
\begin{equation}
\|\dot{\tilde{x}}(t)-\dot{x}(t)\|\leq \big(K(x)\|{\bf F}\|(x)\,+K^2_2(1+D_2(x){\alpha})\big){\alpha}^2\,E^{-1}\,t.
\end{equation}
with ${K}_i$ and $D_2(x)$ functions bounded by constants whose values are unity within an order of magnitude.
\end{teorema}
\section{Comparison of the Solutions of the Vlasov and Averaged
Vlasov Equations}

In this {\it section} we estimate the difference
between the solutions of the Liouville equations associated with the averaged Lorentz connection and
the original Lorentz connection.

Given a non-linear connection characterized by the {\it second order}
 vector field $\chi\in {\bf TTM}$, the associated Liouville equation is $\chi (f)=0$.

{\bf Examples}.
\begin{enumerate}
\item From the connection Lorentz connection $^L\nabla^i_{jk}(x,y)$ [6], one can recover the spray coefficients $^LG(x,y)$, using the homogeneous properties on $y$ and Euler's theorem. In particular, the spray coefficients are:

\begin{displaymath}
^LG^i(x,y)=\,^L\Gamma^i\,_{jk}(x,y)\,y^j y^k=\, \,\Big(\,^{\eta}\Gamma^i\,_{jk} +
\frac{1}{2{\sqrt{\eta(y,y)}}}({\bf F}^i\,_{j}(x)y^m\eta_{mk}+ {\bf
F}^i\,_{k}(x)y^m\eta_{mj})+
\end{displaymath}
\begin{displaymath} +{\bf F}^i\,_m (x)
\frac{y^m}{2\sqrt{\eta(y,y)}}(\eta_{jk}-\frac{1}{\eta(y,y)}\eta_{js}
\eta_{kl}y^s y^l)\Big)y^j y^k=
\end{displaymath}
\begin{displaymath}
=\, \Big(\,^{\eta}\Gamma^i\,_{jk} +
\frac{1}{2{\sqrt{\eta(y,y)}}}\,({\bf F}^i\,_{j}(x)y^m\eta_{mk}+ {\bf
F}^i\,_{k}(x)y^m\eta_{mj})\Big) y^j y^k=
\end{displaymath}
\begin{displaymath}
=\,^{\eta}\Gamma^i\,_{jk}\,y^j y^k +
{{\sqrt{\eta(y,y)}}}\,{\bf F}^i\,_{j}(x)y^j.
\end{displaymath}

From here, one defines the vector field $L\chi$:
\begin{displaymath}
^L\chi=\, y^i\frac{\partial}{\partial x^i}- \,\big( \,^{\eta}\Gamma^i\,_{jk}\,y^j y^k +
{{\sqrt{\eta(y,y)}}}{\bf F}^i\,_{j}(x)y^j\big)\,\frac{\partial}{\partial y^i}.
\end{displaymath}

\item A similar procedure applies to the averaged Lorentz Vlasov vector field. In this case, however, we do not have the simplifications above, except for $\eta(y,y)=1$. Therefore, the spray coefficients are:
  \begin{displaymath}
<\,^LG^i>(x,y)=\,^L\Gamma^i\,_{jk}(x,y)\,y^j y^k=\, \,\Big(\,^{\eta}\Gamma^i\,_{jk} +
<\frac{1}{2{\sqrt{\eta(y,y)}}}({\bf F}^i\,_{j}(x)y^m\eta_{mk}+ {\bf
F}^i\,_{k}(x)y^m\eta_{mj})+
\end{displaymath}
\begin{displaymath} +{\bf F}^i\,_m (x)
\frac{y^m}{2\sqrt{\eta(y,y)}}(\eta_{jk}-\frac{1}{\eta(y,y)}\eta_{js}
\eta_{kl}y^s y^l)\Big)>y^j y^k=
\end{displaymath}
 \begin{displaymath}
 =\,\,\Big(\,^{\eta}\Gamma^i\,_{jk} +
<\frac{1}{2}({\bf F}^i\,_{j}(x)y^m\eta_{mk}+ {\bf
F}^i\,_{k}(x)y^m\eta_{mj})+
\end{displaymath}
\begin{displaymath} +{\bf F}^i\,_m (x)
\frac{y^m}{2}(\eta_{jk}-\eta_{js}
\eta_{kl}y^s y^l)\Big)>y^j y^k=
 \end{displaymath}
\begin{displaymath}
 =\,\,\Big(\,^{\eta}\Gamma^i\,_{jk} +
\frac{1}{2}({\bf F}^i\,_{j}(x)<y^m>\eta_{mk}+ {\bf
F}^i\,_{k}(x)<y^m>\eta_{mj})+
\end{displaymath}
\begin{displaymath} +{\bf F}^i\,_m (x)
(\eta_{jk}<\frac{y^m}{2}>-\eta_{js}
\eta_{kl}\,<\frac{y^m}{2}y^s y^l>)\Big)\,y^j y^k.
 \end{displaymath}
\end{enumerate}
The averaged Vlasov vector field can be written in a similar way as before,
\begin{displaymath}
<^L\chi>=\, y^i\frac{\partial}{\partial x^i}- \,\Big( \,^{\eta}\Gamma^i\,_{jk}\,y^j y^k +\,
\frac{1}{2}({\bf F}^i\,_{j}(x)<y^m>\eta_{mk}+ {\bf
F}^i\,_{k}(x)<y^m>\eta_{mj})+
\end{displaymath}
\begin{displaymath} +{\bf F}^i\,_m (x)
(\eta_{jk}<\frac{y^m}{2}>-\eta_{js}
\eta_{kl}\,<\frac{y^m}{2}y^s y^l>)\Big)\,y^j y^k.
\end{displaymath}

\begin{proposicion}
Let $f$ and $\tilde{f}$ be solutions of the Vlasov equation $^L\chi (f) =0$ and the {\it averaged Vlasov equation} $<^L\chi
> (\tilde{f}) =0$, where $^L\chi$ and $<\,^L\chi
>$ are the spray vector fields obtained from the non-linear connections $^L\nabla$ and $<\, ^L\nabla>$.
Let us assume:
\begin{enumerate}
\item The same hypotheses as those in {\it theorem 2.6},

\item $supp(f)\subset \,supp(\tilde{f})$,

\item $supp(f)$ is a sub-manifold of ${\bf TM}$.
\end{enumerate}
Then for the solutions of the Vlasov and {\it averaged Vlasov's equation} one has the relation:
\begin{displaymath}
|f(t,x(t),\dot{x}(t))-\tilde{f}(t,{x}(t),\dot{{x}}(t))|<\,\big(\tilde{C}(x)\|{\bf F}\|_{\bar{\eta}}
C^2_2(x)(1+B_2(x){\alpha})\big){\alpha}^2\,E^{-2}\,t^2\,+
\end{displaymath}
\begin{equation}
+\big(\tilde{K}(x)\|{\bf F}\|_{\bar{\eta}}(x)\,K^2_2(1+D_2(x){\alpha})\big){\alpha}^2\,E^{-2}\,t.
\end{equation}
for some functions $\tilde{C}(x(t))$ $\tilde{K}(x(t))$ along the geodesic of the Lorentz connection.
\end{proposicion}
{\bf Proof}: $f$ and $\tilde{f}$ are solutions of the
corresponding Vlasov and averaged Vlasov equations respectively. Therefore,  $f$ and
$\tilde{f}$ are constant along the corresponding auto-parallel curves:
\begin{displaymath}
^L\chi_{} f^{}=\frac{d}{dt}f(x(t),\dot{x}(t))=0, \quad
<\,^L\chi_{}>
\tilde{f}^{}=\frac{d}{dt}\tilde{f}(\tilde{x}(t),\dot{\tilde{x}}(t))=0.
\end{displaymath}
For the same initial conditions, the geodesic curves corresponding
to the connections $^L\nabla$ and $<\,^L\nabla>$ are nearby curves
at the instant $t$ in the way described by {\it theorem 2.6}.

Let us introduce the family of {\it interpolating connections},
\begin{displaymath}
^L\nabla_{\epsilon}:=(1-\epsilon)\,^L\nabla +\,\epsilon <\,^L\nabla>.
\end{displaymath}
Each of them has an associated spray vector field
$\chi_{\epsilon},\,\epsilon \in [0,1]$. Therefore, let us consider a solution $f_{\epsilon}(x,y)$ be
the solution of the following Liouville equation:
\begin{displaymath}
^L\chi_{\epsilon} f_{\epsilon}=0.
\end{displaymath}
Since the dependence on $(\epsilon,x,y)$ of the vector field
$^L\chi_{\epsilon} $ is $\mathcal{C}^1$, the solutions of
the differential equation are Lipschitz on the parameter $\epsilon$. We can see this fact in
the following way. The Liouville equation can be written as:
\begin{displaymath}
^L\chi_{\epsilon} f_{\epsilon}=0\,\Leftrightarrow \frac{d}{ds}
f(x_{\epsilon}(s), y_{\epsilon}(s))=0,
\end{displaymath}
where $(x_{\epsilon}(s), y_{\epsilon}(s))$ is an integral curve of the vector
field $^L\chi_{\epsilon}$ restricted to the unit hyperboloid
bundle and such that it is  parameterized by the coordinate time $t$.

One can use standard results from the theory of ordinary differential
equations to study smoothness of the solutions of the above
equation. In particular, the connection coefficients for the
interpolating connection are
\begin{displaymath}
(\,^L\Gamma_{\epsilon})^i\,_{jk}:=(1-\epsilon)\,^L\Gamma^i\,_{jk}\,+\epsilon
<\,^L\Gamma^i\,_{jk}>.
\end{displaymath}
From the formula $(2.3)$ for the coefficients $^L\Gamma^i\,_{jk}$
one can check that $(\,^L\Gamma_{\epsilon})^i\,_{jk}$ are smooth functions in the open set of
time-like vectors $y$ and on the parameter $\epsilon$.

We will give an upper bound on the difference
$|f(t,x(t),\dot{x}(t))-\tilde{f}(t,{x}(t),\dot{{x}}(t))|$.
In order to achieve this,  standard results on
the smoothness of the solution of differential equations are used (see the {\it chapter} 1 of  [14]).
In particular we use that for each $ (\bar{\epsilon}, \bar{x}(s), \bar{y}(s))$, there is
an open neighborhood ${\bf U}_{\bar{\epsilon}}$ of $[0,1]\times { supp(f)}$ containing $ (\bar{\epsilon}, \bar{x}(t), \bar{y}(t))$
such that the solutions of the differential equations are Lipschitz  in ${\bf U}_{\bar{\epsilon}}$:
\begin{displaymath}
|f^{\epsilon}(t,x_{\epsilon}(t),\dot{x}_{\epsilon}(t))-{f}^{\tilde{\epsilon}}(t,{x}_{\tilde{\epsilon}}(t),
\dot{{x}}_{\tilde{\epsilon}}(t))|\leq \,c_1{(\bar{\epsilon}, \bar{x}(t),\dot{\bar{x}}(t)) }\delta((\bar{\epsilon}, \bar{x}(t),\dot{\bar{x}}(t)))\,
+
\end{displaymath}
\begin{displaymath}
+c_2{(\bar{\epsilon}, \bar{x}(t),\dot{\bar{x}}(t)) }\, \|x_{\epsilon}(t)-{x}_{\tilde{\epsilon}}(t)\|_{\bar{\eta}}
+c_3{(\bar{\epsilon}, \bar{x}(t),\dot{\bar{x}}(t)) }\, \|\dot{x}_{\epsilon}(t)-\dot{x}_{\tilde{\epsilon}}(t)\|_{\bar{\eta}}.
\end{displaymath}
$c_i{(\bar{\epsilon}, \bar{x}(t),\dot{\bar{x}}(t)) }$ are constants which depend on the open neighborhood ${\bf U}_{\bar{\epsilon}}$;
$\delta((\bar{\epsilon}, \bar{x}(t),\dot{\bar{x}}(t)))$ is the diameter on the $\epsilon$ component
 where we are applying the Lipschitz condition.

One can always choose a refinement of an open cover of $[0,1]\times supp(f)$ such that both the
 Lipschitz condition, {\it theorem} $(2.6)$ and {\it theorem} $(2.7)$ can be applied simultaneously.
Since $[0,1]$ and $supp(f)$ are compact, we can consider a finite open covering of $[0,1]\times supp(f)$.
 Then, using the above local bound one obtains a global bound of the form:
\begin{displaymath}
|f(t,x(t),\dot{x}(t))-\tilde{f}(t,\tilde{x}(t),\dot{\tilde{x}}(t))|<\,c_1
+c_2\, \|x(t)-\tilde{x}(t)\|_{\bar{\eta}}+c_3\, \|\dot{x}(t)-\dot{\tilde{x}}(t)\|_{\bar{\eta}}.
\end{displaymath}
The constants $c_i$ are
finite (by definition of Liptschitz and by compactness of the
interval $[0,1]$). The functions $f$ and $\tilde{f}$ are constant
along the respective geodesics when parameterized by the corresponding proper time. Therefore,
\begin{displaymath}
|f(t,x(t),\dot{x}(t))-\tilde{f}(t,\tilde{x}(t),\dot{\tilde{x}}(t))|=\,
 |f(0,x(0),\dot{x}(0))-\tilde{f}(0,\tilde{x}(0),\dot{\tilde{x}}(0))|.
\end{displaymath}
Let us assume the same initial conditions $x(0)=\tilde{x}(0)$ and $\dot{x}(0)=\dot{\tilde{x}}(0)$.
Since the difference $|f(t,x(t),\dot{x}(t))-\tilde{f}(t,{x}(t),\dot{{x}}(t))|$ is a smooth function of
 $\|x(t)-\tilde{x}(t)\|_{\bar{\eta}}$ and $\|\dot{x}(t)-\dot{\tilde{x}}(t)\|_{\bar{\eta}}$, one has that
\begin{displaymath}
0\leq \,c_1\leq \,n\cdot \bar{K}_1\|x(t)-\tilde{x}(t)\|_{\bar{\eta}}+\,  \bar{K}_1\|\dot{x}(t)-\dot{\tilde{x}}(t)\|_{\bar{\eta}}
\end{displaymath}
for some constants $K_i$.
Then we have that:
\begin{displaymath}
|f(t,x(t),\dot{x}(t))-\tilde{f}(t,{x}(t),\dot{{x}}(t))|\leq
\end{displaymath}
\begin{displaymath}
\leq \,|f(t,x(t),\dot{x}(t))-\tilde{f}(t,\tilde{x}(t),\dot{\tilde{x}}(t))|+\,
|\tilde{f}(t,x(t),\dot{x}(t))-\tilde{f}(t,\tilde{x}(t),\dot{\tilde{x}}(t))|.
\end{displaymath}
The first term is bounded by $c_1$, which is bounded by $\,n\cdot \bar{K}_1\|x(t)-\tilde{x}(t)\|_{\bar{\eta}}+\,
\cdot \bar{K}_1\|\dot{x}(t)-\dot{\tilde{x}}(t)\|_{\bar{\eta}}$. The second term can be developed in Taylor series on the differences $\|x(t)-\tilde{x}(t)\|_{\bar{\eta}}$
 and $\|\dot{x}(t)-\dot{\tilde{x}}(t)\|_{\bar{\eta}}$, since $\tilde{f}$ is smooth. Therefore:
\begin{displaymath}
|f(t,x(t),\dot{x}(t))-\tilde{f}(t,{x}(t),\dot{{x}}(t))|\leq\,\big(\tilde{C}(x)\|{\bf F}\|\,
{C}^2_2(1+B_2(x){\alpha})\big){\alpha}^2\,E^{-2}\, t^2\,+
\end{displaymath}
\begin{displaymath}
+\big(\tilde{K}(x)\|{\bf F}\|(x)\,{K}^2_2(1+{D}_2(x){\alpha})\big){\alpha}^2\,E^{-1}\,t.
\end{displaymath}\hfill$\Box$

\section{The Charged Cold Fluid Model from the Averaged Vlasov
Model}

\begin{definicion} Given a semi-Randers space $({\bf M}, \eta, [A])$,
the averaged Maxwell-Vlasov model is defined by the dynamical
variables $({\bf F},\tilde{f})$ determined by the coupled system of equations:
\begin{equation}
 <\, ^L \chi> \tilde{f}=0,
\quad
\tilde{V}=\int_{supp(\tilde{f}_x)}\, dvol(x,y) y \tilde{f}(x,y),\quad \rho=\int_{
supp(\tilde{f}_x)} \,dvol(x,y)\tilde{f}(x,y).
\end{equation}
\end{definicion}
In this case, the remaining dynamical variable $f(x,y)$ is determined by an integral-differential equation:
\begin{equation}
 <\, ^L \chi>\tilde{f}=0,
\end{equation}
where $^L\chi$ is the Liouville vector field of the averaged
Lorentz dynamics associated with the external electromagnetic
field ${\bf F}$.

Since we will use the results of {\it section 3} and reference [6], the distribution
 function $\tilde{f}$ is at least of type $\mathcal{C}^1$. Since the support is compact, there are several
 Sobolev norms [13] which are defined. We will write our results in terms of these norms.

There is one question which deserves to be clarified. Firstly, we have assumed that the $supp(\tilde{f}_x)$ is
 a smooth manifold. In principle, this is not necessarily the case. However, we have restricted
 to the case where $supp(f_x)\subset supp(\tilde{f}_x)$. This can be achieved in the following way.
\begin{proposicion}
Let $<\,^L\chi>\tilde{f}(x,y)=0$ and $^L\chi f(x,y)=0$ be such that
\begin{enumerate}

\item The domain of definition of the vector $<\, L^\chi>$ is an open sub-manifold of ${\bf \Sigma}$.

\item $supp(f_x)\subset supp(\tilde{f}_x)$ for all $x\in {\bf M}$,

 \item $supp(f_x)$ and $\partial(supp(f_x))$ are sub-manifolds of ${\bf \Sigma}_x$ for all $x\in {\bf M}$.
 \end{enumerate}
Then one can reduce $supp(\tilde{f}_x)\longrightarrow supp(f_x)$ for all $x\in {\bf M}$.
\end{proposicion}
{\bf Proof}:
Let us consider the product of functions $\tilde{f}g$, where $<\,^L\chi>\tilde{f}=0$ and
 the function $g$ is a $\mathcal{C}^{\infty}$ {\it bumpy} function defined on the domain of the vector field $<\,^L\chi>$.
 Since both, the domain of definition of the $\Omega(<\,^L\chi>)$ and $supp(f_x)$ are sub-manifolds, this function exists [15]. Therefore,
\begin{displaymath}
g(x,y)=0, \quad (x,y)\in supp(\tilde{f})\setminus supp({f}_x),\quad g(x,y)=0,\quad (x,y)\in \partial{supp({f}_x)}
\end{displaymath}
and all the derivatives are zero on $\partial supp({f}_x)$. Then, one can perform the following calculation:
\begin{displaymath}
<\,^L\chi>(\tilde{f}g)=\,g<\,^L\chi>\tilde{f}\,+\tilde{f}<\,^L\nabla>g=0.
\end{displaymath}
Therefore, one can always restrict the solutions of $<\,^L\chi>\tilde{f}=0$ in such a way that $supp(\tilde{f}_x)=supp(f_x)$.\hfill$\Box$

Also note that using $\mathcal{C}^{\infty}$ bumpy functions we can obtain a distribution function $\tilde{f}$
which is bounded by $1$: since $supp(\tilde{f})$ is compact, we can consider the positive function
$\frac{1}{\|\tilde{f}\|_{\infty}}\,\tilde{f}$, with $<\,^L\chi>\,\tilde{f}=0$.
Then, $<\,^L\chi>\,\frac{1}{\|\tilde{f}\|_{\infty}}\,\tilde{f}=0$ and the new solution is bounded by $1$ (here $\|\tilde{f}\|_{\infty}:=\,max\{f(x,y),\quad y\in\, supp(f)\}$).

In the calculations below we assume that $supp(\tilde{f}_x)$ is orientable,
with a volume form $dvol(x,y)$.
If this is not the case, it follows from {\it proposition 4.2} that we can
 approximate $({ \tilde{f}_x},supp(\tilde{f}_x))$ by
  $(f_x, supp({f}_x))$  and then use $dvol(x,y)$ induced from the
  volume form on the unit hyperboloid ${\bf \Sigma}_x$ when we calculate the averages.

 Using equation (3.1), it follows that
 the error induced by the substitution $\tilde{f}\longrightarrow f$ is of order $\alpha^2$ and due to {\it proposition 4.2}
 is not harmful reducing the support such that $supp(\tilde{f}_x)\longrightarrow supp(f_x)$.
 Therefore, in the following calculations, when it is convenient we can use $dvol(x,y)$ as a measure and substitute $supp(\tilde{f}_x)$ by $supp(f_x)$.
 However we will write the calculation assuming that $supp(\tilde{f})$ is a manifold with a volume form $dvol(x,y)$,
 since this is a more general statement.

Let ${\bf F}$ be a closed differential $2$-form defining the Liouville vector field $^L\chi$.
 Let us consider the Sobolev spaces $(\mathcal{W}^{1,1}(supp(\tilde{f}_x)),\|\cdot\|_{1,1})$
  and $(\mathcal{W}^{0,2}(supp(\tilde{f}_x)),\,\|\cdot\|_{0,2})$ [13].
   Recall that the space of smooth functions $\mathcal{F}(supp(\tilde{f}_x))$ is a
   sub-space of both of those Sobolev spaces (for compact $supp(\tilde{f}_x)$).

Let us denote by $\delta(x,y)=<y>(x)-y$.
\begin{teorema}
Let ${\bf M}$ be an $n$-dimensional space-time manifold and $<\,^L\chi>$ the vector
 field associated with the averaged Lorentz force equation. Assume that:
\begin{enumerate}
\item The distribution function is such that $\tilde{f}_x,\, \partial_j \tilde{f}\in
 \mathcal{F}(supp(\tilde{f}_x))\, \subset\mathcal{W}_{1,1}(supp(\tilde{f}_x))$,

\item The function  $\delta(x,y), \partial_j \delta(x,y)\in \mathcal{W}_{0,2}(supp(\tilde{f}_x))$.

\end{enumerate}
Then
\begin{equation}
\|<\, ^LD>_{\tilde{V}} \tilde{V}(x)\|_{\bar{\eta}}\leq \,\frac{vol^{\frac{1}{2}}_E(supp(\tilde{f}_x))}{vol(supp(\tilde{f}_x))}
\,(\sum_{k}\,\|\partial_0\, log(\delta^k_x)\|_{0,2})\,\,
\cdot \|\tilde{f}_x\|_{1,1}\cdot {\alpha}^2\,+ O(\alpha^3).
\end{equation}
where $\delta_x(\cdot):=\delta(x,\cdot)$ and
\begin{displaymath}
\tilde{V}^i(x)=<\hat{y}^i>_{\tilde{f}}(x)\,:=\frac{1}{\int_{supp(\tilde{f})_x} f(x,y)
 dvol(x,{y})}\,\int_{supp(\tilde{f})_x}\,dvol(x,y)\,f(x,y) y^i.
\end{displaymath}
The volumes are the following
\begin{displaymath}
vol(supp(\tilde{f}_x)):=vol({\bf \Sigma}_x);\quad vol_E(supp(\tilde{f}_x)):=\int_{supp(\tilde{f}_x)}\, dvol(x,{y});
\end{displaymath}
the derivative in equation $(4.3)$ refers to the local frame such that the vector $U=(U_0,\vec{o})$
\end{teorema}
{\bf Proof:} Because the averaged Lorentz connection is an affine
connection on {\bf M}, given a point $x\in {\bf M}$, there is a coordinate system where the
connection coefficients are zero at that point,  $<\, ^L\Gamma>^i _{jk}=0$.
Therefore, for any given point $x\in {\bf M}$, one can choose a {\it normal coordinate system}
such that the following relation holds:
\begin{equation}
y^j \partial_j \tilde{f}(x,y)=0
\end{equation}
at that point. The existence of normal coordinates is of fundamental importance in the sequel.
 For instance, one can get an expression for the covariant derivative of $\tilde{V}$ along the integral curve of $\tilde{V}$,
\begin{equation}
<\, ^LD>_{\tilde{V}} \tilde{V}=(\tilde{V}^j\partial_j \tilde{V}^k )\frac{\partial}{\partial
x^k},
\end{equation}
using a coordinate frame $\{\frac{\partial}{\partial
x^k},\,\,k=0,...,n-1\}$. Note that this expression is not a partial differential equation
because it holds only at the point $x$.

Developing the above expression further, we have the relation:
\begin{displaymath}
<\, ^LD>_{\tilde{V}} \tilde{V}(x)=\frac{1}{vol(supp(\tilde{f}_x))}\int_{supp({f}_x)}\,dvol(x,y)\,
y^j \tilde{f}(x,{y})
\cdot
\end{displaymath}
\begin{displaymath}
\cdot \partial_j \big(\frac{1}{vol(supp(\tilde{f}_x))}\,\int_{supp(\tilde{f}_x)}dvol(x,\hat{y}) \hat{y}^k
\tilde{f}(x,\hat{y})\big).
\end{displaymath}
$supp(\tilde{f}_x)$ is compact and small compared with $E(x)$, for each $x\in {\bf M}$.
It is the right hand of this equation which we should estimate,
\begin{displaymath}
\frac{1}{vol(supp(\tilde{f}_x))}\,\Big(\int_{supp(f_x)}\,dvol(x,y)\, y^j \tilde{f}(x,{y})
\partial_j \big(\frac{1}{vol(supp(\tilde{f}_x))}\,\int_{supp(\tilde{f}_x)}\,dvol(x,\hat{y}) \hat{y}^k
\tilde{f}(x,\hat{y})\big)\,\Big)=
\end{displaymath}
\begin{displaymath}
=\frac{1}{vol(supp(\tilde{f}_x))}\Big( \,\int_{supp(\tilde{f}_x)}\,dvol(x,y)\, y^j
\tilde{f}(x,y)\,\big(-\,\frac{\int_{supp(\tilde{f}_x)}\, dvol(x,\hat{y})\,
\tilde{f}(x,\hat{y})\,\hat{y}^k}{vol^2( supp(\tilde{f}_x))}\cdot
\end{displaymath}
\begin{displaymath}
\cdot\partial_j\,(\int_{supp(\tilde{f}_x)}\, dvol(x,\tilde{y})\, f(x,\tilde{y}))\big)\Big)
+\frac{1}{vol(supp(\tilde{f}_x))}\,\Big(\, \int_{supp(\tilde{f}_x)} \,dvol(x,y)\,y^j\,
\tilde{f}(x,y)\,\cdot
\end{displaymath}
\begin{displaymath}
\cdot\partial_j\,\big(\, \int_{supp(\tilde{f}_x)}\,dvol(x,\hat{y})\,
\hat{y}^k\,\tilde{f}(x,\hat{y})\,\big)\Big)=
\end{displaymath}
\begin{displaymath}
=-\frac{1}{vol^2(supp(\tilde{f}_x))}\Big(\,\int_{supp(\tilde{f}_x)}\,dvol(x,y)\, y^j \tilde{f}(x,y)\,
\partial_j\,\big(\,\int_{supp(\tilde{f}_x)}\, dvol(x,\hat{y})\,\tilde{f}(x,\hat{y})\,\,\big)\,<y^k>\,\Big)\,+
\end{displaymath}
\begin{displaymath}
+\frac{1}{vol^2(supp(\tilde{f}_x))}\,\Big(\,\int_{supp(\tilde{f}_x)}\, dvol(x,y)\,y^j\,
\tilde{f}(x,y)\,\partial_j\,(\int_{supp(\tilde{f}_x)}\, dvol(x,\hat{y})\,\hat{y}^k\,
\tilde{f}(x,\hat{y})\,\big)\,\Big).
\end{displaymath}
Shifting the variable of integration $-\hat{y}^k+<\hat{y}^k>=-\delta^k (x,\hat{y})$, one obtains for the above expression
\begin{displaymath}
<\, ^LD>_{\tilde{V}} \tilde{V}(x)=-\frac{1}{vol^2(supp(\tilde{f}_x))}\Big(\,\int_{supp(\tilde{f}_x)}\,dvol(x,y)\, y^j
\tilde{f}(x,y)\cdot
\end{displaymath}
\begin{displaymath}
\partial_j\,\big(\,\int_{supp(\tilde{f}_x)}\, dvol\hat{y}\,
\tilde{f}(x,\hat{y})\,\,\big)\,<y^k>\,\Big)\,+
\end{displaymath}
\begin{displaymath}
+\frac{1}{vol^2(supp(\tilde{f}_x))}\,\Big(\,\int_{supp(\tilde{f}_x)}\, dvol(x,y)\,y^j\,
\tilde{f}(x,y)\,\partial_j\,(\int_{supp(\tilde{f}_x)}\, dvol(x,\hat{y})\,(<y^k>+\delta^k(x,\hat{y}))\,
\tilde{f}(x,\hat{y})\,\big)\,\Big)=
\end{displaymath}
\begin{displaymath}
=\frac{1}{vol^2(supp(\tilde{f}_x))}\,\Big(\,\int_{supp(\tilde{f}_x)}\, dvol(x,y)\,y^j\,
\tilde{f}(x,y)\,\partial_j\,(\int_{supp(\tilde{f}_x)}\, dvol(x,\hat{y})\,\delta^k(x,\hat{y})\,
\tilde{f}(x,\hat{y})\,\big)\,\Big).
\end{displaymath}
Since $y^i\partial_i f(x,y)=0$ at the origin $x$ of a normal coordinate system and since $\tilde{f}_x$ is a smooth function of $y$, we can Taylor expanded the above expression:
\begin{displaymath}
\frac{1}{vol^2(supp(\tilde{f}_x))}\,\Big(\,\int_{supp(\tilde{f}_x)}\, dvol(x,y)\,y^j\,
\tilde{f}(x,y)\,\partial_j\,(\int_{supp(\tilde{f}_x)}\, dvol(x,\hat{y})\,\delta^k(x,\hat{y})\,
\big(\tilde{f}(x,{y})+\frac{\partial f}{\partial \hat{y}^l}\, (\hat{y}^l-y^l)\,\big)\,\Big).
\end{displaymath}
Since the averaged connection is an affine connection, there is a coordinate system such that $ <\,^L\Gamma>^i_{jk} =0$.
 This is reflected in the expression for the averaged Vlasov equation, which has the form $y^j\,\partial _j
\tilde{f}(x,y)=0$ at one given point $x\in {\bf M}$. Then, we get the following for the above expression
\begin{displaymath}
\Big(<\, ^LD>_{\tilde{V}} \tilde{V}(x)\Big)^k=\frac{1}{vol^2(supp(\tilde{f}_x))}\,\Big(\,\int_{supp(\tilde{f}_x)}\, dvol(x,y)\,y^j\,
\tilde{f}(x,y)\,\cdot
\end{displaymath}
\begin{displaymath}
\cdot \partial_j\,(\int_{supp(\tilde{f}_x)}\, dvol(x,\hat{y})\,\delta^k(x,\hat{y})\,
\frac{\partial f}{\partial \tilde{y}^l}\, (\hat{y}^l-y^l)\,\big)\,\Big).
\end{displaymath}
$(\hat{y}^l-y^l)$ and ${\delta}^k(x,y)$ are bounded by the
diameter ${\alpha}$ (remember that in taking the moments we have substituted
$\tilde{f}_x,\,  supp(\tilde{f}_x)$ by $(f_x,\, supp(f_x))$. Therefore,
\begin{displaymath}
\Big\|<\, ^LD>_{\tilde{V}}
\tilde{V}\Big\|_{\bar{\eta}}\,=
\end{displaymath}
\begin{displaymath}
=\frac{1}{vol^2(supp(\tilde{f}_x))}\,\Big\|\Big(\,\int_{supp(\tilde{f}_x)}\,
dvol(x,y)\,y^j\,\tilde{f}(x,y)\,
\cdot
\end{displaymath}
\begin{displaymath}
\cdot \partial_j\,(\int_{supp(\tilde{f}_x)}\,
dvol(x,\hat{y})\,\delta^k(x,\hat{y})\,\frac{\partial f}{\partial
\hat{y}^l}\, (\hat{y}^l-y^l)\partial_k\,\big)\,\Big)\Big\|_{\bar{\eta}}\,\leq
\end{displaymath}
\begin{displaymath}
\leq\frac{1}{vol^2(supp(\tilde{f}_x))}\,\Big|\Big(\,\int_{supp(\tilde{f}_x)}\,
dvol(x,y)\,y^j\,\tilde{f}(x,y)\,\partial_j\,\frac{\partial f}{\partial
\tilde{y}^l}\,\Big)\Big|\,
\cdot
\end{displaymath}
\begin{displaymath}
\cdot\Big\|\Big(\int_{supp(\tilde{f}_x)}\,
dvol(x,\hat{y})\,\delta^k(x,\hat{y})\,\,
(\hat{y}^l-y^l)\partial_k\,\Big)\,\Big\|_{\bar{\eta}}+
\end{displaymath}
\begin{displaymath}
+\frac{1}{vol^2(supp(f_x))}\,\Big|\Big(\,\int_{supp(\tilde{f}_x)}\,
dvol(x,y)\,y^j\,\tilde{f}(x,y)\,\frac{\partial f}{\partial
\tilde{y}^l}\,\Big)\Big|\,
\cdot
\end{displaymath}
\begin{displaymath}
\cdot \Big\|\Big(\int_{supp(\tilde{f}_x)}\,
dvol(x,\hat{y})\,\,\partial_j\delta^k(x,\hat{y})\,\,
(\hat{y}^l-y^l)\partial_k\,\Big)\,\Big\|_{\bar{\eta}}\,\leq
\end{displaymath}
\begin{displaymath}
\leq \frac{1}{vol^2(supp(\tilde{f}_x))}\,\Big|\Big(\,\int_{supp(\tilde{f}_x)}\,
 dvol(x,y)\,y^j\,\tilde{f}(x,y)\,\partial_j\,\frac{\partial
f}{\partial \tilde{y}^l}\,\Big)\Big|\,
\cdot
\end{displaymath}
\begin{displaymath}
\cdot \Big(\int_{supp(\tilde{f}_x)}\,
dvol(x,\hat{y})\,\big\|\delta^k(x,\hat{y})\,\,
(\hat{y}^l-y^l)\partial_k\Big\|_{\bar{\eta}}\,\Big)+
\end{displaymath}
\begin{displaymath}
+\frac{1}{vol^2(supp(\tilde{f}_x))}\,\Big|\Big(\,\int_{supp(\tilde{f}_x)}\, dvol(x,y)\,y^j\,\tilde{f}(x,y)\,\frac{\partial
f}{\partial \tilde{y}^l}\,\Big)\Big|\,
\cdot
\end{displaymath}
\begin{displaymath}
\cdot \Big(\int_{supp(\tilde{f}_x)}\,
dvol(x,\hat{y})\,\partial_j\big\|\delta^k(x,\hat{y})\,\,
(\hat{y}^l-y^l)\partial_k\big\|_{\bar{\eta}}\,\Big).
\end{displaymath}
One can find a bound for each of these integrals.
For instance, using the Hoelder inequality for integrals [13]
\begin{displaymath}
\Big|\int_{{\bf X}} \psi \phi\, \,d\mu\,\Big|\leq\,\Big(\int_{{\bf X}}
\big|\psi\big|^{p}\,\Big)^{1/p} \Big(\int_{{\bf X}}  \big|\phi\big|^{q}\,\Big)^{1/q},\,\,\,
\frac{1}{p}+\frac{1}{q}=1,\,\,1\leq p,q\leq\infty.
\end{displaymath}
We will use this inequality several times for the case $p=q=2$.
Then, we obtain the following:
\begin{displaymath}
\Big\|\Big(\int_{supp(\tilde{f}_x)}\, dvol(x,\tilde{y})\,\,
(\tilde{y}^l-y^l)\,\delta^k(x,y)\partial_k\,\Big\|_{\bar{\eta}}\,\Big)\leq \Big(\int_{supp(\tilde{f}_x)}\, dvol(x,\hat{y})\,
\big|(\hat{y}^l-y^l)\big|\,\big\|\,
\,\delta^k(x,\hat{y})\partial_k\,\big\|_{\bar{\eta}}\,\Big)\leq
\end{displaymath}
\begin{displaymath}
\leq \,\Big(\int_{supp(\tilde{f}_x)}\, dvol(x,\hat{y})\big|\,
(\hat{y}^l-y^l)\big|^2\,\Big)^{\frac{1}{2}}
\cdot\, \Big(\int_{supp(\tilde{f}_x)}\, dvol(x,\hat{y})\,
\big\|\,\delta^k(x,\hat{y})\partial_k\,\big\|^2_{\bar{\eta}}\,\Big)^{\frac{1}{2}}\leq
\end{displaymath}
\begin{displaymath}
 \leq \, vol^{\frac{1}{2}}_E(supp(\tilde{f}_x))\,\alpha \cdot
  \Big(\int_{supp(\tilde{f}_x)}\, dvol(x,\hat{y})\,\big\|\,\delta^k(x,\hat{y})\partial_k\,\big\|^2_{\bar{\eta}}\Big)^{\frac{1}{2}}.
\end{displaymath}
The Euclidean volume was
\begin{displaymath}
vol_E(supp(\tilde{f}_x)):=\int_{supp(\tilde{f}_x)}\, dvol(x,{y}).
\end{displaymath}
To bound the second factor we use the following [6]:
\begin{displaymath}
\|\delta(x,y)\|_{\bar{\eta}}\,\leq\|<\hat{y}>(x)\,-y\|_{\bar{\eta}}\,\leq \|\epsilon+\hat{y}-y\|_{\bar{\eta}} \,
\leq \|\epsilon\|_{\bar{\eta}}+\,\|\hat{y}-y\|_{\bar{\eta}}\,\leq \frac{1}{2}\,\alpha\,+\alpha=\frac{3}{2} \alpha.
\end{displaymath}
The parameter $\epsilon$ is related to the norm of the mean velocity and $\hat{y}$
 is in the support of the distribution $f$. Therefore, a bound on the integral is
\begin{displaymath}
\Big\|\Big(\int_{supp(\tilde{f}_x)}\, dvol(x,\tilde{y})\,
(\tilde{y}^l-y^l)\,\delta^k(x,\tilde{y})\partial_k\,\Big)\Big\|_{\bar{\eta}}\,\,\leq \frac{3}{2}\cdot vol_E(supp(\tilde{f}_x))\cdot \alpha^2.
\end{displaymath}
Similarly one obtains the following bound:
\begin{displaymath}
\Big|\Big(\int_{supp(\tilde{f}_x)}\, dvol(x,\hat{y})\,\partial_j (\delta^k(x,y))\,
(\hat{y}^l-y^l)\partial_k\,\Big)\Big\|_{\bar{\eta}}\,\leq \, \Big(\int_{supp(\tilde{f}_x)}\, dvol(x,\hat{y})\,\big|
(\hat{y}^l-y^l)\big|^2\,\Big)^{\frac{1}{2}}\cdot \,
\end{displaymath}
\begin{displaymath}
\cdot\,\Big(\int_{supp(\tilde{f}_x)}\, dvol(x,{y})\,
\big\|\,\partial_j (\delta^k(x,{y}))\partial_k\big\|^2_{\bar{\eta}}\,\Big)^{\frac{1}{2}}\,\leq
\end{displaymath}
\begin{displaymath}
\leq \, vol_E^{\frac{1}{2}}(supp(\tilde{f}_x))\,\alpha
\cdot \, \Big(\int_{supp(\tilde{f}_x)}\, dvol(x,{y})\,\big\|\,
\partial_j (\delta^k(x,y))\partial_k\big\|^2 _{\bar{\eta}}\,\Big)^{\frac{1}{2}}.
\end{displaymath}
Because of the definition of the corresponding Sobolev norm one obtains
\begin{displaymath}
 \Big(\int_{supp(\tilde{f}_x)}\, dvol(x,{y})\,\big\|\partial_j (\delta^k(x,{y}))\,
(\tilde{y}^l-y^l)\partial_k\big\|_{\bar{\eta}}\,\Big)\leq \,vol^{\frac{1}{2}}_E(supp(\tilde{f}_x))
\,{\alpha}\cdot\|\partial_j \delta^k_x\|_{0,2}\,=
\end{displaymath}
\begin{displaymath}
=\,vol^{\frac{1}{2}}_E(supp(\tilde{f}_x))
\,{\alpha}\cdot\sum_k\,\|\delta^k_x\partial_j log(\delta^k_x)\|_{0,2}
\leq \,vol^{\frac{1}{2}}_E(supp(\tilde{f}_x))
\,{\alpha}^2\cdot\sum_k\,\|\partial_j log(\delta^k_x)\|_{0,2}.
\end{displaymath}
Similarly,
\begin{displaymath}
\Big|\Big(\,\int_{supp(\tilde{f}_x)}\, dvol(x,y)\,y^j\,\tilde{f}(x,y)\,\partial_j\,\frac{\partial
f}{\partial \tilde{y}^l}\,\Big)\Big|\,\leq \,\sum^{n-1}_{j=0}\,\Big(\,\int_{supp(\tilde{f}_x)}\, dvol(x,y)\,
\big|y^j\,\tilde{f}(x,y)\big|^2\,\Big)^{\frac{1}{2}}\cdot
\end{displaymath}
\begin{displaymath}
\cdot\Big(\,\int_{supp(\tilde{f}_x)}\, dvol(x,y)\,\big|\partial_j\,\frac{\partial \tilde{f}(x,y)}{\partial y^k}\big|^2\,\Big)^{\frac{1}{2}}.
\end{displaymath}
The second factor is equal to the Sobolev norm $\|\partial_j \tilde{f}_x\|_{1,1}$. The first factor is bounded in the following way:
\begin{displaymath}
\Big(\,\int_{supp(\tilde{f}_x)}\, dvol(x,y)\,\big|y^j\,\tilde{f}(x,y)\big|^2\,\Big)^{\frac{1}{2}}\, \leq
 \Big(\,\int_{supp(\tilde{f}_x)}\, dvol(x,y)\,|y^j|^2\,\tilde{f}(x,y)\,\Big)^{\frac{1}{2}}=
 \end{displaymath}
 \begin{displaymath}
= vol(supp(\tilde{f}))\,(<|y^j|^2>)^{\frac{1}{2}}.
\end{displaymath}
Therefore, we get the bound:
\begin{displaymath}
\Big|\Big(\,\int_{supp(\tilde{f}_x)}\, dvol(x,y)\,y^j\,\tilde{f}(x,y)\,\partial_j\,\frac{\partial
f}{\partial \tilde{y}^l}\,\Big)\Big|\,\leq vol(supp(\tilde{f}_x))\,\Big(\sum^{n-1}_{j=0}(<|y^j|^2>)^{\frac{1}{2}}\cdot\, \|\partial_j \,\Big) \tilde{f}_x\|_{1,1}.
\end{displaymath}
In a local frame where the vector field  $<U>=(U^0,\vec{0})$, this contraction can be re-written as
\begin{displaymath}
(<(y^0)^2>)^{\frac{1}{2}}\cdot\, \|\partial_0 \tilde{f}_x\|_{1,1}=
\|(<|y^0|^2>)^{\frac{1}{2}}\cdot\, \partial_0 \tilde{f}_x\|_{1,1}=\|(<(y^0)^2>)^{\frac{1}{2}}\cdot\, \partial_0 \tilde{f}_x\|_{1,1}=
\end{displaymath}
\begin{displaymath}
=\|(<(y^0)^2\cdot\, (\partial_0 \tilde{f}_x)^2>)^{\frac{1}{2}}\|_{1,1}=
\|(<(y^j\cdot\, \partial_j \tilde{f}_x)^2>)^{\frac{1}{2}}\|_{1,1}.
\end{displaymath}
Using normal coordinates system associated with the affine connection $<\,^L\nabla>$ and using the averaged Vlasov equation, one obtains that $\|(<(y^j\cdot\, \partial_j \tilde{f}_x)^2>)^{\frac{1}{2}}\|_{1,1}=0$.

Finally, we can bound the following integral
\begin{displaymath}
\Big|\Big(\,\int_{supp(\tilde{f}_x)}\, dvol(x,y)\,y^j\,\tilde{f}(x,y)\,\frac{\partial
f}{\partial {y}^l}\,\Big)\Big|\,\leq \Big(\,\int_{supp(\tilde{f}_x)}\, dvol(x,y)|\,y^j\,\tilde{f}(x,y)\,\frac{\partial
f}{\partial {y}^l}|\,\Big)\,\leq
\end{displaymath}
\begin{displaymath}
\leq \,\sum^{n-1}_{j=0}\,\Big(\,\int_{supp(\tilde{f}_x)}\, dvol(x,y)\,|y^j\,\tilde{f}(x,y))|^2\,\Big)^{\frac{1}{2}}\
\cdot(\,\int_{supp(\tilde{f}_x)}\, dvol(x,y)\,|\frac{\partial \tilde{f}(x,y)}{\partial y^k}|^{\frac{1}{2}}\Big)^{\frac{1}{2}}.
\end{displaymath}
Like in the previous integral, we get:
\begin{displaymath}
|\Big(\,\int_{supp(\tilde{f}_x)}\, dvol(x,y)\,y^j\,\tilde{f}(x,y)\,\frac{\partial
f}{\partial {y}^l}\,\Big)|\leq vol(supp(\tilde{f}_x))\,\cdot\, (<|y^j|^2>)^{\frac{1}{2}}\cdot\, \|\tilde{f}_x\|_{1,1}.
\end{displaymath}

Using these bounds, we obtain the following relation:
\begin{displaymath}
\|<\, ^LD>_{\tilde{V}} \tilde{V}(x)\|_{\bar{\eta}}\leq \frac{1}{vol^2(supp(\tilde{f}_x))}\,\Big|\Big(\,\int_{supp(\tilde{f}_x)}\,
dvol(x,y)\,y^j\,\tilde{f}(x,y)\,\partial_j\,\frac{\partial f}{\partial
{y}^l}\,\Big)\Big|\,\cdot\,
\end{displaymath}
\begin{displaymath}
\cdot \frac{3}{2}\,vol_E(supp(\tilde{f}_x))\,{\alpha}^2+
\end{displaymath}
\begin{displaymath}
+\,\frac{1}{vol^2(supp(\tilde{f}_x))}\,\sum^{n-1}_{j=0}\Big|\Big(\,\int_{supp(\tilde{f}_x)}\,
dvol(x,y)\,y^j\,\tilde{f}(x,y)\,\frac{\partial f}{\partial
\tilde{y}^l}\,\Big)\Big|\cdot
\end{displaymath}
\begin{displaymath}
\cdot \,vol^{\frac{1}{2}}_E(supp(\tilde{f}_x))
\cdot {\alpha}^2\cdot(\sum_k\,\|\partial_j\, log(\delta^k_x)\|_{0,2})\,\leq
\end{displaymath}
\begin{displaymath}
\leq\frac{1}{vol^2(supp(\tilde{f}_x))}\, \cdot vol(supp(\tilde{f}_x))\,\cdot\Big(\sum^{n-1}_{j=0} \, (<|y^j|^2>)^{\frac{1}{2}}\,
\cdot \|\tilde{f}_x\|_{1,1}\cdot\,\,vol^{\frac{1}{2}}_E(supp(\tilde{f}_x))\cdot
\end{displaymath}
\begin{displaymath}
\cdot {\alpha}^2\cdot(\sum_k\,\|\partial_j\, log(\delta^k_x)\|_{0,2})\Big)\,=
\end{displaymath}
\begin{displaymath}
=\frac{vol^{\frac{1}{2}}_E(supp(\tilde{f}_x))}{vol(supp(\tilde{f}_x))}\,\cdot\Big(\,\sum^{n-1}_{j=0}(<|y^j|^2>)^{\frac{1}{2}}
\,(\sum_k\,\|\partial_j\, log(\delta^k_x)\|_{0,2})\big)\,
\cdot \|\tilde{f}_x\|_{1,1}\cdot {\alpha}^2.
\end{displaymath}
In a local frame where the vector field $U$ has components $(U^0,\vec{0})$, one has
\begin{displaymath}
\sum^{n-1}_{j=0} (<|y^j|^2>)^{\frac{1}{2}}
\,(\sum_k\,\|\partial_j\, log(\delta^k_x)\,\|_{0,2}=(<|y^0|^2>)^{\frac{1}{2}}
\,(\sum_k\,\|\partial_0\, log(\delta^k_x)\,\|_{0,2}=
\end{displaymath}
\begin{displaymath}
=\,\sum_k\,\|<(y^0)^2>\, \partial_0\, log(\delta^k_x)\,\|_{0,2}.
\end{displaymath}
Note that the normal coordinate system (that we are using) coincides with the adapted coordinate system,
 associated to the vector field $U=(U^0,\vec{0}$. In this coordinate system, there is a bound $<y^0>\leq 1+\alpha$. Therefore,
 \begin{displaymath}
 \sum^{n-1}_{j=0}\,(<|y^j|^2>)^{\frac{1}{2}}
\,(\sum_k\,\|\partial_j\, log(\delta^k_x)\,\|_{0,2}\leq (<|y^0|^2>)^{\frac{1}{2}}\cdot \,(\sum_k\,\|\partial_0\, log(\delta^k_x)\,\|_{0,2}\,\cdot \,(1+\alpha).
 \end{displaymath}
We get the following result,
 \begin{displaymath}
 \|<\, ^LD>_{\tilde{V}} \tilde{V}(x)\|_{\bar{\eta}}\leq \,\frac{vol^{\frac{1}{2}}_E(supp(\tilde{f}_x))}{vol(supp(\tilde{f}_x))}
\,(\sum_{k}\,\|\partial_0\, log(\delta^k_x)\|_{0,2})\,\,
\cdot \|\tilde{f}_x\|_{1,1}\cdot {\alpha}^2\,+ O(\alpha^3).
 \end{displaymath}\hfill$\Box$

 Usually we are interested in the behavior of the fluid in compact regions of the space-time {\bf M}:
 \begin{corolario}
 For compact domains ${\bf K}\subset {\bf M}$ and under the same hypotheses as in {\it theorem 4.3}, the following relation holds:
 \begin{displaymath}
 \|<\, ^LD>_{\tilde{V}} \tilde{V}(x)\|_{\bar{\eta}}\leq \,\cdot C({\bf K})\cdot \alpha^2\,  +O(\alpha^3),
 \end{displaymath}
 for some constant $C({\bf K})$.
 \end{corolario}
 {\bf Proof}: Take for the constant $C({\bf K})$ to be
 \begin{displaymath}
 C({\bf K})=max_{x\in {\bf K}} \Big\{\,\frac{vol^{\frac{1}{2}}_E(supp(\tilde{f}_x))}{vol(supp(\tilde{f}_x))}
\,(\sum_{k}\,\|\partial_0\, log(\delta^k_x)\|_{0,2})\,\,
\cdot \|\tilde{f}_x\|_{1,1}\,\Big\}.
 \end{displaymath}
 \hfill$\Box$

These bounds are quite general are meaningful in the case that $1>> \alpha$.

In the preceding results the ultra-relativistic limit ($E>>1$) was not essential, although
it in intrinsically used when we write conditions on the smoothness conditions of $supp(\tilde{f})$ and the measure used.

\paragraph{}
Let us consider the normalized averaged velocity vector field:
\begin{displaymath}
\tilde{u}=\frac{\tilde{V}}{\eta(\tilde{V},\tilde{V})^{1/2}}.
\end{displaymath}
Since $<\,^LD>$ does not preserve the Minkowski metric $\eta$,
the covariant derivative of
$\tilde{u}$ in the
direction of $\tilde{u}$ using the Lorentz connection $^LD$ is
\begin{equation}
<\, ^LD>_{\tilde{u}} \tilde{u} = \frac{1}{\eta(\tilde{V},\tilde{V})}\,
<\, ^LD>_{\tilde{V}} \tilde{V} \,
+\frac{1}{2}\Big(\tilde{V}\cdot\big(log(\eta(\tilde{V},\tilde{V}))\big) \Big)\tilde{V}.
\end{equation}
The first terms is bounded by {\it theorem 4.3}, since $\eta(\tilde{V},\tilde{V})>1$.
The total derivative of $\eta(\tilde{V},\tilde{V})$ along a trajectory of $\tilde{V}$
is
\begin{displaymath}
\frac{d}{dt}\big(\eta(\tilde{V},\tilde{V})\big)=\mathcal{L}_{\tilde{V}}
\big(\eta(\tilde{V},\tilde{V})\big)=\tilde{V}\cdot\big(\eta(\tilde{V},\tilde{V})\big)
=2\eta\Big(<\, ^LD>_{\tilde{V}} \tilde{V} ,\tilde{V}\Big)+\Big(<\, ^LD>_{\tilde{V}} \eta
\Big) (\tilde{V},\tilde{V}).
\end{displaymath}
As we have proved the first term is of order $\alpha^2$.
Using normal coordinates for $<\, ^L\nabla>$, one can compute the
second term,
\begin{displaymath}
\big(<\, ^LD>_{\tilde{V}} \eta\,\big) (\tilde{V},\tilde{V}) =
\eta(\tilde{V},\tilde{V}) {\bf F}_{\mu m} <\delta^m(x,y)\,
\delta^s(x,y)\, \delta^l(x,y)> \tilde{V}^{\mu}\tilde{V}_s\tilde{V}_l.
\end{displaymath}
We can estimate these contributions:
\begin{proposicion}
Under the same assumptions as in {\it theorem 4.3}, the following relation holds:
\begin{equation}
<\,^LD>_{\tilde{u}} \tilde{u}\leq \,\frac{vol^{\frac{1}{2}}_E(supp(\tilde{f}_x))}{vol(supp(\tilde{f}_x))}
\,(\sum_{k}\,\|\partial_0\, log(\delta^k_x)\|_{0,2})\,\,
\cdot \|\tilde{f}_x\|_{1,1}\cdot {\alpha}^2\,+ O(\alpha^3),
\end{equation}
where $n$ is the dimension of ${\bf M}$.
\end{proposicion}
{\bf Proof}: The first term of the right hand of the equation ${\it 4.6}$ is bounded by {\it theorem 4.3}.
 The second term is bounded using Hoelder's inequality for integrals in the following
 (this is a particular case of [16, pg 62]):
 \begin{displaymath}
 \Big|\int_S dvol(z)\, f_1(z)\cdot\cdot\cdot\,f_m(z)\Big|\,\leq \prod^{m}_{k=1}\,
 \Big(\int_S dvol(z)\, \big|f_k(z)\big|^{p_k}\Big)^{\frac{1}{p_k}},\,\,\,\, \sum_{k} p_k =1,\,\, 1\leq p_k\leq \infty.
 \end{displaymath}
In particular one can apply this inequality to the third order
moment $<\delta^m(x,y)\,\delta^s(x,y)\, \delta^l(x,y)>$:
\begin{displaymath}
\Big|<\delta^m(x,y)\,
\delta^s(x,y)\, \delta^l(x,y)>\Big|\,=\,\frac{1}{vol(supp(\tilde{f}_x))}\,
\Big|\int_{supp(\tilde{f}_x)} dvol(x,{y})\,f(x,y)\,\delta^m(x,y)\,
\delta^s(x,y)\, \delta^l(x,y)\Big|\,
\end{displaymath}
\begin{displaymath}
\leq \frac{1}{vol(supp(\tilde{f}_x))}\,
\Big(\int_{supp(\tilde{f}_x)} dvol(x,y)\,|\tilde{f}(x,y)\,\delta^m(x,y)|^3\,\Big)^{\frac{1}{3}}\cdot
\Big(\int_{supp(\tilde{f}_x)} dvol(x,y)\,|\tilde{f}(x,y)\delta^s(x,y)|^3\,\Big)^{\frac{1}{3}}\cdot
\end{displaymath}
\begin{displaymath}
\cdot\Big(\inf_{supp(\tilde{f}_x)} dvol(x,y)\,|\tilde{f}(x,y)\delta^l(x,y)|^3\,\Big)^{\frac{1}{3}}.
\end{displaymath}
Since the distribution function is positive on $supp(\tilde{f})$ and $\tilde{f}_x\leq 1$.
Then, since equation $(3.1)$, one can substitute in the integrations $\tilde{f}_x\longrightarrow f_x$,
which implies that
\begin{displaymath}
\Big|<\delta^m(x,y)\,
\delta^s(x,y)\, \delta^l(x,y)>\Big|\,=O(\alpha^3).
\end{displaymath}
 Since the norm $\bar{\eta}(<y>,<y>)\ge
1$, one gets a third degree monomial term
in ${\alpha}$ for the covariant derivative $<\,
^LD>_{\tilde{u}} \tilde{u} $.\hfill$\Box$

\begin{corolario}
Under the same assumptions as in {\it theorem 4.3}, in a compact domain ${\bf K}\subset {\bf M}$
 one obtains the following asymptotic relation:
\begin{equation}
<\,^LD>_{\tilde{u}} \tilde{u}\leq \, \tilde{C}({\bf K})\cdot \alpha^2 \,+O(\alpha^3),
\end{equation}
for a convenient constant $\tilde{C}({\bf K})$.
\end{corolario}
\paragraph{}
Let us consider a local {\it Lorentz congruence}, which is a set of
auto-parallel curves of the Lorentz connection $^LD$, for a set
of initial conditions at each $(t_0,\vec{x})$, $\vec{x}\in{\bf
M}_{t_0}$ where ${\bf M}_{t_0}\hookrightarrow {\bf M}$ is a $3$-dimensional
spatial sub-manifold. One can consider in a similar way the congruence
associated with the averaged Lorentz connection with the same initial conditions.
Note that, while the Lorentz connection preserves the Lorentz norm
  $\eta(\dot{x},\dot{x})$ of the tangent vectors of the geodesics,
  this is not the case for the averaged Lorentz connection).

\begin{teorema}
Let {\bf F} be a closed $2$-form and $^L\nabla$
the associated non-linear Lorentz connection. Then the solutions of
 the Lorentz force equation $^{\eta}\nabla_{\dot{x}} {\dot{x}}= (\iota _{\dot{x}} {\bf F})^\sharp$
can be approximated by the integral curves of the normalized mean velocity field $u$ of the
distribution function $f(x,y)$, where $f(x,y)$ is a solution of the associated Vlasov equation
$^L\chi f=0$. The difference is controlled by polynomial functions at least of order $2$ in ${\alpha}$:
\begin{equation}
\|\,^LD_{{u}} {u}\|\,\leq \,{a}_2(x)\, {\alpha}^2\,+O(\alpha^3).
\end{equation}
\end{teorema}
{\bf Proof:} We repeat an argument that we have used before.  By {\it proposition 3.1}, both distribution
functions $f$ and $\tilde{f}$, solutions of $^L\chi f=0$ and $<\,^L\chi>\tilde{f}$, are such that:
\begin{displaymath}
|f(t,x(t),\dot{x}(t))-\tilde{f}(t,{x}(t),\dot{{x}}(t))|\leq\,\big(\tilde{C}(x)\|{\bf F}\|(x)
C^2_2(x)(1+B_2(x){\alpha})\big){\alpha}^2\,E^{-2}\,t^2\,+
\end{displaymath}
\begin{displaymath}
+\big(\tilde{K}(x)\|{\bf F}\|(x)\,K^2_2(1+D_2(x){\alpha})\big){\alpha}^2\,E^{-1}\,t.
\end{displaymath}
Therefore, the corresponding mean velocity fields  are nearby as well, because the linearity of the averaging operation and because of the relation:
\begin{displaymath}
|f(t,x(t),\dot{x}(t))|-\,|\tilde{f}(t,{x}(t),\dot{{x}}(t))|\,\leq |f(t,x(t),\dot{x}(t))-\tilde{f}(t,{x}(t),\dot{{x}}(t))|\,\leq
\end{displaymath}
\begin{displaymath}
\leq \,\big(\tilde{C}(x)\|{\bf F}\|(x)
C^2_2(x)(1+B_2(x){\alpha})\big){\alpha}^2\,E^{-2}\,t^2\,
+\big(\tilde{K}(x)\|{\bf F}\|(x)\,K^2_2(1+D_2(x){\alpha})\big){\alpha}^2\,E^{-1}\,t.
\end{displaymath}
Then, their corresponding
integral curves and the associated local congruences are also numerically similar. By {\it corollary 4.6},
 for narrow distributions,
 the normalized mean field $\tilde{u}$ associated with $\tilde{f}$,
\begin{displaymath}
\|<\,^LD>_{\tilde{u}} \tilde{u}\|\,\leq\, \tilde{a}_2\, {\alpha}^2\,+\,O({\alpha}^3).
\end{displaymath}
for some known function $\tilde{a}_2(x)$. Remember that, one can interpolate smoothly between the connections $^LD$ and $<\,^LD>$.
 Therefore, locally, one can interpolate smoothly between their integral curves.
 Also, because the smoothness of the solutions of the geodesic equations in the parameter of interpolation, there is a function ${a}$
 in a small open neighborhood of {\bf M} such that
\begin{displaymath}
^LD_{{u}} {u}\,\leq \,{a}_2(x)\, {\alpha}^2\,+O(\alpha^3).
\end{displaymath}
\hfill$\Box$

\section{Discussion}

{\it Theorem 4.5} and {\it Corollary 4.6} show when the charged cold fluid
model in the description of the dynamics of a collection of
particles interacting with an external electromagnetic field in the ultra-relativistic regime.
It is particularly interesting that we have proved the result without using
additional hypothesis on the behavior of the higher moments of the distribution, except that the distribution is
 narrow and some convenient smoothness conditions.

There are some technical issues that we would like to mention.
\begin{enumerate}

\item The distribution functions $\tilde{f}$ and $f$ are at least $\mathcal{C}^1$. However, let us consider a Dirac
 delta distribution with support invariant by the flow of the Lorentz force [6]:
\begin{equation}
f(x,y)=\Psi(x)\,\delta(y-V(x)).
\end{equation}
Since the width of the distribution is zero, ${\alpha}=0$. One can use this distribution as a solution
 of the Vlasov equation in two space-time dimension spaces, for a proper value of the function $\Psi$ .  This example and the fact that the bounds found in {\it section 4}
  are formulated using Sobolev norms suggests the possibility to generalize the results to bigger spaces.

\item The need for the condition $supp(f_x)\subset supp(\tilde{f}_x)$ needs to be investigated. We have seen that one can reduce
$supp(\tilde{f}_x)$ conveniently. Indeed, the inclusion is non-trivial, since the delta function
 limit is a solution of both, the averaged Vlasov equation and the original Vlasov equation.
On the other hand, the integro differential equation $<\,^L\chi>\,\tilde{f}=0$
      is in some sense {\it simpler} than the original Vlasov equation $^L\chi (f)=0$. This is
       because the averaged connection is {\it simpler} than the Lorentz connection. This fact provides an insight why
        $supp({f_x})\subset supp(\tilde{f}_x)$.

\item The same method can be applied to other fluid equations. Therefore, depending on the specific bounds, one can decide which model is better in a particular situation

\item Although the results presented in this work are Lorentz covariant
 (in the sense that the normal coordinate system has an invariant definition)
, they are written in a non-covariant way. This is because we have introduced
  norms associated with particular reference frames. These norms are essential in our approach. At the moment,
   we don't know how to give a Lorentz invariant presentation of the content.
    Therefore, further research is necessary on this direction.

\end{enumerate}

\paragraph{}
\paragraph{}
{\bf Acknowledgement.}

I would like to thank to Dr. D. Burton and Dr. Volker Perlick for many discussions about the content presented in this work.

\end{document}